\documentclass[aps,prl,reprint,superscriptaddress]{revtex4-2}

\usepackage{float}
\usepackage{stmaryrd}
\usepackage{gensymb}
\usepackage{textcomp}
\usepackage{graphicx}
\usepackage{amsmath}
\usepackage{amssymb}
\usepackage[colorinlistoftodos]{todonotes}
\usepackage{siunitx}
\usepackage[absolute,overlay]{textpos}
\usepackage[colorlinks=true, citecolor={black}, urlcolor={blue!60!black}, linkcolor = {black}]{hyperref}
\usepackage{comment}
\usepackage{bold-extra}
\usepackage[normalem]{ulem}
\usepackage{comment}

\makeatletter

\newcommand{\fmarki}{*}
\newcommand{\fmarkii}{\ensuremath{\dagger}}
\newcommand{\fmarkiii}{\ensuremath{\ddagger}}
\newcommand{\fmarkiv}{\ensuremath{\mathsection}}
\newcommand{\fmarkv}{\ensuremath{\mathparagraph}}
\newcommand{\fmarkvi}{\ensuremath{\|}}
\newcommand{\fmarkvii}{**}
\newcommand{\fmarkviii}{\ensuremath{\dagger\dagger}}
\newcommand{\fmarkix}{\ensuremath{\ddagger\ddagger}}
\def\@fnsymbol#1{{\ifcase#1\or \fmarki\or \fmarkii\or \fmarkiii\or \fmarkiv\or \fmarkv\or \fmarkvi\or \fmarkvii\or \fmarkviii\or \fmarkix \else\@ctrerr\fi}}
\makeatother

\renewcommand{\fmarki}{\ensuremath{\dagger}}
\renewcommand{\fmarkii}{*}

\def\uppi~{$\mathrm{\pi}$}

\newcommand{\fl}[1]{\textbf{(#1)}}

\newcommand{\var}[2]{$#1_{\mathrm{#2}}$}

\begin{document}

\title{Gate reflectometry in a minimal Kitaev chain device }

\author{Yining Zhang}
\affiliation{QuTech and Kavli Institute of Nanoscience, Delft University of Technology, Delft, 2600 GA, The Netherlands}

\author{Ivan Kulesh}
\affiliation{QuTech and Kavli Institute of Nanoscience, Delft University of Technology, Delft, 2600 GA, The Netherlands}

\author{Sebastiaan L. D. ten Haaf}
\affiliation{QuTech and Kavli Institute of Nanoscience, Delft University of Technology, Delft, 2600 GA, The Netherlands}

\author{Nick van Loo}
\affiliation{QuTech and Kavli Institute of Nanoscience, Delft University of Technology, Delft, 2600 GA, The Netherlands}

\author{Francesco Zatelli}
\affiliation{QuTech and Kavli Institute of Nanoscience, Delft University of Technology, Delft, 2600 GA, The Netherlands}

\author{Tijl Degroote}
\affiliation{QuTech and Kavli Institute of Nanoscience, Delft University of Technology, Delft, 2600 GA, The Netherlands}

\author{Christian G. Prosko}
\affiliation{QuTech and Kavli Institute of Nanoscience, Delft University of Technology, Delft, 2600 GA, The Netherlands}

\author{Srijit Goswami}\email{s.goswami@tudelft.nl}
\affiliation{QuTech and Kavli Institute of Nanoscience, Delft University of Technology, Delft, 2600 GA, The Netherlands}

\begin{abstract}
Hybrid quantum dot (QD)-superconductor system can be used to realize Majorana zero modes in artificial Kitaev chains. These chains provide a promising platform for the realization of Majorana qubits.
Radio-frequency (RF) gate reflectometry is a fast, non-invasive, and sensitive technique that can be used to read out such qubits.  
In this work, we use gate reflectometry to probe two QDs coupled via a semiconductor-superconductor hybrid segment. 
We demonstrate that gate sensing can resolve charge stability diagrams and clearly distinguish between elastic cotunneling and crossed-Andreev reflection, the two key processes that allow one to form a Kitaev chain.
Furthermore, we show that this information is accessible, even when the system is completely decoupled from the from the normal leads. In this closed regime, we show that the observed quantum capacitance signal is indicative of parity switching between the even and odd ground states.
Our measurements in both open and closed regimes confirm that gate reflectometry captures the essential features of interdot coupling and parity dynamics. 
\end{abstract}

\maketitle

\section{Introduction}
Hybrid systems with quantum dots coupled to superconductors~\cite{sau2012realizing,leijnse2012parity,tsintzis2022creating} allow for the engineering Majorana zero modes (MZMs) in artificial Kitaev chains~\cite{kitaev2001unpaired}. 
With the recent realization of minimal Kitaev chains~\cite{dvir2023realization,ten2024two} there now exist concrete proposals to use such systems to implement Majorana based qubits~\cite{leijnse2012parity,szechenyi2020parity,tsintzis2023roadmap}, and to demonstrate non-abelian operations~\cite{liu2023fusion,boross2024braiding,tsintzis2024majorana}. 
In order to to read out such qubits, fast measurement techniques are crucial.
Radio-frequency (RF) reflectometry has been demonstrated as a powerful readout technique, enabling high-speed, non-invasive detection of quantum dot charge and spin states by monitoring impedance changes in resonator circuits~\cite{esterli2019small, vigneau2023probing}.
RF sensing has been widely used to sense charge~\cite{urdampilleta2015charge, de2019rapid, reilly2007fast, barthel2010fast,prosko2024flux} and spin~\cite{fedele2021simultaneous, west2019gate, betz2015dispersively} in QD arrays. 
More recently, RF reflectometry has been applied to QD-superconductor hybrid systems to sense Cooper pair splitting~\cite{de2023controllable,van2019revealing} and quasiparticle poisoning (QPP)~\cite{menard2019suppressing, nguyen2023electrostatic,quantum2025interferometric, van2025single}. 
Gate-based RF reflectometry is particularly well suited for probing Kitaev chains, as it integrates naturally with quantum dots—the essential building blocks of these hybrid systems—enabling compact, non-invasive access to probe the system.

In this work, we use gate reflectometry techniques to study a minimal Kitaev chain device in a two-dimensional electron gas (2DEG) with two quantum dots coupled by a semiconductor-superconductor hybrid segment. 
We perform experiments in two measurement configurations: with or without normal reservoirs. 
When the system is coupled to normal leads, we can measure charge stability diagrams (CSDs) with both lead reflectometry and gate reflectometry. 
Gate reflectometry can clearly distinguish the type of coupling between QDs, consistent with lead reflectometry measurements.
When the system is decoupled from the normal leads, gate reflectometry can still be used to probe the quantum capacitance of the system and we observed quantum capacitance from even and odd ground states transitions.
The observation of both even and odd parity in  the same measurement indicates that the ground state of the system can switch by quasiparticle poisoning.
We verify this by partially coupling the normal leads, allowing us to artificially inject quasiparticles into the system.

\section{methods}

\begin{figure}
\centering
    \includegraphics[width = 0.48\textwidth]{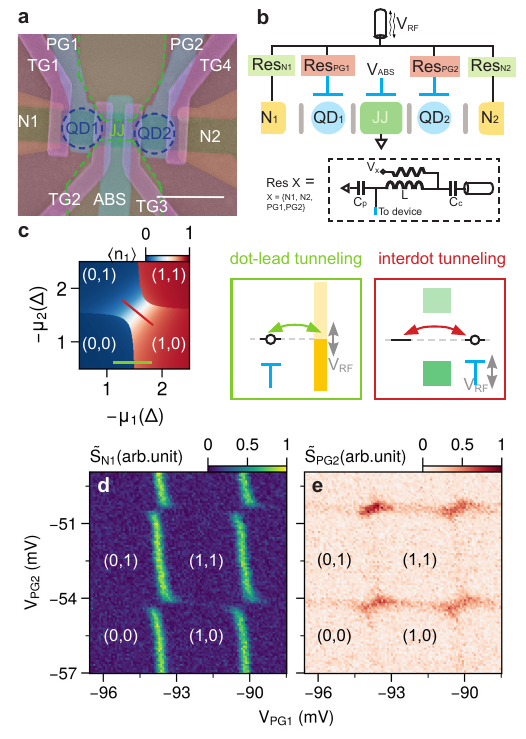}
    \caption{\textbf{Experimental setup and charge stability diagrams.} 
    \fl{a} False-color SEM image of the device with labeled electrostatic gates and metallic leads. The two Al leads outlined with dashed green lines in Fig.~1a are connected by a loop (see Fig.~S1a). The scale bar represents \SI{500}{\nano\meter}. 
    \fl{b} Schematic of the RF circuit. The bottom panel shows a schematic of each resonator, consisting of a parasitic capacitance \var{C}{p}, an inductor \var{L}{}, a coupling capacitor \var{C}{c}, and a resistor as the bias tee. 
    \fl{c} Left: simulated charge occupation number on QD1. Sharp contours separate ground states with different parity (e.g., the green mark), while within a single parity sector, the charge occupation changes smoothly (e.g., the red mark). Middle and right: schematic energy diagrams for the tunneling processes corresponding to the two markers.
    \fl{d, e} Charge stability diagrams measured from normal lead N1 and on plunger gate PG2, respectively. 
   }
    \label{fig:Fig1}
\end{figure}
Fig.~1a shows a false-color scanning electron microscopic image of a device that is lithographically identical to the one measured.
The device is fabricated on an InSbAs-Al 2DEG~\cite{moehle2021insbas}. 
There are three layers of gates: the first layer (red) defines the one-dimensional channel; the second layer (blue) independently controls the chemical potential of each section; and the third layer (pink) defines the tunnel barriers. 
Each gate layer is separated by \SI{20}{\nano m} dielectric $\mathrm{Al}_2\mathrm{O}_3$.
The two QDs are coupled by a Josephson junction (JJ) that hosts Andreev bound states. 
The junction is formed by two aluminum tips (colored green with dashed outline). 
The full device SEM image and basic characterization of the QDs and JJ can be found in the supplementary material Fig.~S1.

A resonator chip is used to perform multiplexed RF sensing~\cite{hornibrook2014frequency}.  
The schematic of the RF circuit is shown in Fig.~1b. 
There are 4 resonators used in this experiment: two of them connect to the two normal leads, and two of them connect to the two QD plunger gates.
The lower panel of Fig.~1b shows a schematic of an individual resonator with a bias-tee that allows the application of DC bias or gate voltage to the device.
The reflected complex voltage \var{S}{11} is recorded for each resonator. 
For better visualization, the signal is projected onto a line that maximizes the signal magnitude, and normalized to a single real value \var{\Tilde{S}}{}. 
The detailed signal analysis method is described in Fig.~S4.  

When resonant tunneling occurs between normal leads and the device or between the two QDs, the impedance of the device changes. 
The impedance change results in changes in \var{S}{11}. 
More specifically, the resistive component of the device changes the magnitude of the resonator response, while the capacitive component shifts the resonator frequency~\cite{vigneau2023probing}.
In low-dimensional quantum systems, like QDs, charge tunneling causes a small correction term of the total capacitance of the gate electrode due to the finite density of states in the system~\cite{vigneau2023probing,ashoori1992single,petersson2010charge}.
This is known as quantum capacitance, given by $C_q=e^2\frac{dN}{d\mu}$
where $N$ is the charge expectation and $\mu$ is the chemical potential. 
Notably, the main contribution to our signal—quantum capacitance—arises only when charge states are coherently hybridized~\cite{mizuta2017quantum,esterli2019small}.\newline

To understand the quantum capacitance signal of a DQD system better, a calculation of the charge expectation number of a QD in a DQD system is shown in Fig.~1c on the left. 
At the bottom left corner and the top right corner, the ground state parity of DQD is even, i.e., (0,0) or (1,1). 
A sharp contour distinguishes the even and odd ground states.
At the charge degeneracy points where two charge states (0,1) and (1,0) start to hybridize (red marker in Fig.~1c left panel), electrons can hop between the two QD levels (shown schematically in Fig.~1c right panel). 
This interdot hopping results in a change in quantum capacitance change (since it is proportional to $d\langle N_1\rangle /d\mu_1$),  and is detectable in the gate sensing signal. On the other hand, when the charge occupation number of the ground state changes, as indicated by the green marker, electrons can tunnel from the normal lead to the DQD, thereby switching the parity of DQD (schematic in middle panel of Fig.~1c). 
This process will result in a strong response on the lead resonator.

\section{Results}
\subsection{Using gate sensing to probe the interdot coupling}

It has been well established that transport measurements of CSDs can reveal the type of coupling between QDs~\cite{wang2022singlet,wang2023triplet,zatelli2024robust}.
To verify that gate sensing provides the same information, we simultaneously measure a CSD with gate reflectometry on PG2 (Fig.~1d) and lead reflectometry via N1 (Fig.~1e). 
The tunnel barrier between QD2 and the normal lead (TG4) is pinched off, while the outer barrier for QD1 (TG1) is configured to the tunneling regime.
Sweeping the plunger gates of the two QDs, we can measure tunneling processes in different sections of the device. 
When a dot level aligns with the Fermi level, dot-lead tunneling occurs, which results in a strong signal on the lead resonator. 
This process corresponds to the signal shown in Fig.~1d. 
Near joint charge degeneracy points, the two charge states in the two QDs start to hybridize, indicated by the avoided crossings in the CSDs.
The resonator response only show QD1 coulomb resonanses because QD2 is decoupled from the normal lead N2 by pinching off TG4 barrier. 
When we use gate sensing to read out the same CSD, the resonator instead responds strongly when the system reaches a joint charge degeneracy point. 
As discussed earlier, a quantum capacitance signal appears when charge distribution changes adiabatically. 
We observe the resonator response exactly at the avoided crossings in Fig.~1d, indicating interdot tunneling. 
When the dot levels are far away from the degeneracy point or in Coulomb blockade, the tunneling is suppressed and gate sensing gives no response. 
Therefore, we confirm that gate sensing provides complementary information about the system, consistent with RF reflectometry measurements from the normal lead.

Having shown that our setup is capable of measuring CSDs with gate sensing, we proceed with probing different kinds of couplings between the two QDs. 
The presence of the superconductor and Andreev bound states (ABSs) allows for tunable elastic cotunneling (ECT) and cross Andreev reflection (CAR) between the QDs~\cite{wang2022singlet,liu2022tunable,wang2023triplet}, which is instrumental for the  realization of artificial Kitaev chains~\cite{dvir2023realization,ten2024two}.
Detecting the dominant type of coupling is important for tuning up these Kitaev chains. The type of interdot coupling can be controlled by changing the chemical potential of the hybrid region.
ECT occurs when coherent hybridization between the dot levels in the odd parity manifold, an electron can tunnel from one dot to the other mediated by the ABS.
In the CSD measured by lead reflectometry, an antidiagonal avoided crossing is shown in Fig.~2b with the (0,1) and (1,0) region connected, which is the signature of the ECT coupled regime. 
The same measurement with gate reflectometry has a strong signal in the diagonal direction in Fig.~2c, corresponding to the interdot tunneling near the charge degeneracy point. 
By sweeping the \var{V}{ABS} gate voltage, we can tune the interdot coupling from ECT-like to CAR-like. 
A schematic of the CAR process is shown in Fig.~2d. 
When the CAR coupling is dominant and two quantum dot levels are antisymmetric along the Fermi level, two electrons can form or split a Cooper pair mediated by Andreev bound states~\cite{wang2022singlet,liu2022tunable,wang2023triplet}.
In the CSD shown in Fig.~2e, the direction of the avoided crossing changed from from antidiagonal to diagonal, connecting the (0,0) and (1,1) in the CSD. 
Near the charge degeneracy point, a strong gate sensing signal arises along the antidiagonal direction.
Therefore, from the direction of the gate sensing signal we can infer the type of coupling between the two QDs.

\begin{figure}
\centering
    \includegraphics[width = 0.48\textwidth]{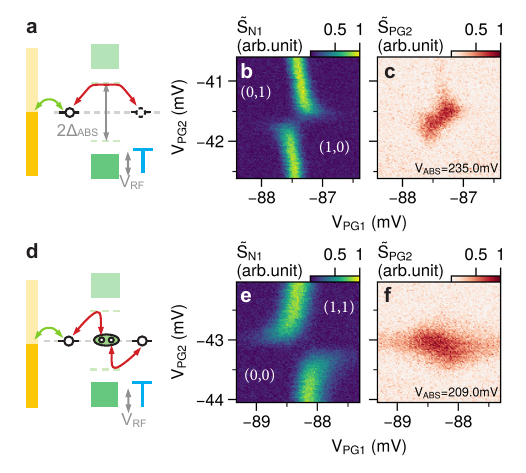}
    \caption{\textbf{Gate sensing of the interdot couplings.} 
    \fl{a} Schematic of ECT coupling The green dashed lines represent the ABS energy levels. The red arrow indicates the interdot hopping mediated by the ABS. 
    \fl{b,c} CSDs with normalized reflected RF voltage \var{\Tilde{S}}{} on N1 (\textbf{b}) and PG2 (\textbf{c}) in the ECT-dominated regime.
    \fl{d} Schematic of CAR coupling.
    \fl{e,f} CSD with normalized reflected RF voltage \var{\Tilde{S}}{} on N1 (\textbf{e}) and PG2 (\textbf{f}) in the CAR-dominated regime. The coupling type can be tuned by sweeping the ABS gate voltage \var{V}{ABS}. 
   }
    \label{fig:Fig2}
\end{figure}

\subsection{Gate sensing in a closed double dot system}
\begin{figure}
\centering
    \includegraphics[width = 0.48\textwidth]{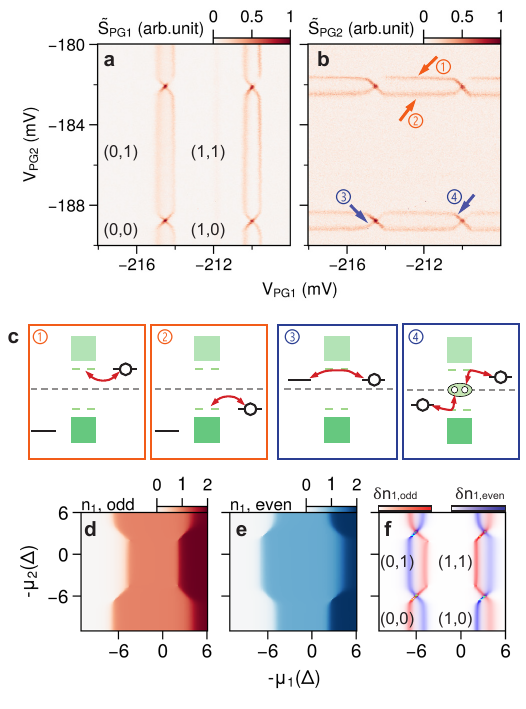}
    \caption{\textbf{Gate reflectometry CSDs of a closed system.} 
    \fl{a,b} CSDs measured by gate reflectometry on the plunger gates of both quantum dots. Blue and orange arrows indicate different tunneling processes, as illustrated in \textbf{(c)}.
    \fl{c} Schematics of the possible tunneling processes labeled in \textbf{(b)}. Processes \textcircled{1} and \textcircled{2} correspond to QD-ABS tunneling when one dot is off resonance and the other dot level aligned with the ABS energy, Process \textcircled{3} and \textcircled{4} correspond to the interdot tunneling events when both dot levels are below the ABS energy (ECT and CAR).
    \fl{d,e} Numerical simulation of charge expectation number for the odd and even ground states of one dot in a DQD coupled by ABS.
    \fl{f} Numerical derivative of the ground state charge with respect to the chemical potential of QD1, $\mu_1$. The signal qualitatively represents quantum capacitance ($C_q$) measured via gate reflectometry. The plot has contributions of $C_q$ from both the odd and even ground states.
   }
    \label{fig:Fig3}
\end{figure}
A minimal Majorana parity qubit device comprises of two pairs of the DQD, and their joint parity (either even or odd) can be used as the basis of the qubit~\cite{leijnse2012parity}.
When implementing a parity qubit, it is necessary to decouple the system from the leads to prevent relaxation and quasiparticle poisoning. 
In this configuration, while lead reflectometry is no longer available, gate sensing can still be used to probe the system. 
To demosntarte this, we pinch off both tunnel barriers (TG1 and TG4) to decouple the system from normal reservoirs, and study the system via gate sensing.

The alternating RF voltages on the plunger gates (PG1 and PG2) allow for electron tunneling between the two dots or between a QD and the ABS, resulting in the quantum capacitance signals.
We find that the CSDs of a closed system measured by gate sensing (Fig.~3a,b) look significantly different to the open system (Fig~1e). 
We observe Coulomb blockade (white background) separated by dot resonance signals. 
The relative charge occupation is labeled in Fig.~3a. 
Instead of one strong Coulomb peak when the QD is on resonance (as shown in Fig.~1d), we observe two parallel lines, as indicated by the orange arrows in Fig.~3b. 
Near the joint charge degeneracy points, we observed cross-like signals, indicated by the blue arrows in Fig.~3b. 

In order to understand the microscopic origin of these features we schematically outline the relevant tunneling processes (Fig.~3c). 
When a QD level is aligned with the ABS level and the other dot is off resonance, electrons can hop between the QD and the ABS, as illustrated in Fig.3c \textcircled{1} and \textcircled{2}. 
Similar tunneling process can also occur directly between the two QDs when they are aligned or antialigned within the ABS level. In these cases, the QDs hybridize due to ECT (Fig.~3c \textcircled{3}) or CAR (Fig.~3c \textcircled{4}) coupling mediated by the ABS.
The corresponding features present themselves near the joint charge degeneracy points, shown as a cross-like pattern.   

To get a better understanding of the double line feature, we compare our experiments with a simple numerical simulation of the quantum capacitance for a DQD coupled by ABS~\cite{tsintzis2022creating, zatelli2024robust} (detailed description in supplementary). We begin with considering the even and odd ground states separately, and calculate the charge occupation of each QD in odd and even ground states as the function of chemical potential of the two dots.
The charge occupation of QD1 at odd and even ground state are plotted in Fig.~3d,e, respectively. 
As we sweep the chemical potential of QD1, the occupation changes from 0 to 2 with distinct steps. 
The qualitative representation of quantum capacitance signal can be determined by taking derivative of $\langle N \rangle$. The derivative of charge of odd state and even state are plotted in Fig.~3f with red and blue, respectively.

If the DQD is coupled to normal leads, the system can always relax to the lowest energy ground state by exchanging electrons with the reservoirs in order to flip the parity when necessary. 
However, when the DQD is decoupled from normal reservoirs, ideally the system maintains a fixed parity, because no electrons can tunnel from the normal lead. 
If this were the case, we should have observed the signal corresponding to only one parity state.
Instead, we observe the quantum capacitance signal corresponding to both the even and odd state, which indicates parity flipping within the integration time of each data point.
We attribute this parity flipping mechanism to quasiparticle poisoning (QPP)~\cite{aumentado2004nonequilibrium, albrecht2017transport, olivares2014dynamics}. 
The source of this QPP could either be the superconducting leads where Cooper pairs are broken by some high energy processes, or from the normal leads, due to a residual coupling to the DQD.

\subsection{Poisoning from the normal reservoir}
\begin{figure}
\centering
    \includegraphics[width = 0.48\textwidth]{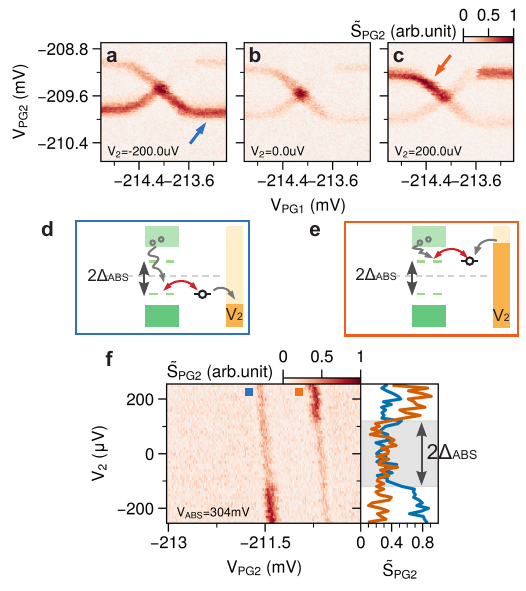}
    \caption{\textbf{Effect of the normal lead.} 
    \fl{a-c} CSDs measured by gate reflectometry on QD2 under different bias voltages \var{V}{2} (annotated in each panel). The arrows in \textbf{(a)} and \textbf{(c)} indicate the enhanced tunneling processes.
    \fl{d,e} Schematic energy diagrams illustrating possible poisoning events originating from the superconductor and the normal lead, corresponding to the arrows in \textbf{a} and \textbf{c}, respectively. 
    \fl{f} Gate reflectometry signal on QD2 as a function of \var{V}{2}, with QD1 off resonance. The right panel shows the amplitude of \var{\Tilde{S}}{PG2} along the dashed line in the main plot. The weaker portion of the signal corresponds to the ABS energy gap.
   }
    \label{fig:Fig4}
\end{figure}
In order to study the effect of the normal leads on the poisoning, we slightly reduce the barrier to N2 (by modifying TG4), while ensuring that the measured current remains below the noise floor of our current meter.
When the bias voltage is kept at 0, the CSD remains similar to the case where the normal lead is fully decoupled, as shown in Fig.~4b.
In this case, quantum capacitance signals from both even and odd states show comparable strength.
By applying a bias voltage to N2, we artificially induce poisoning in the system.
We measure the CSD with \SI{-200}{\micro V} bias voltage applied on N2, shown in Fig.~4a.
Compared to Fig.~4b, the lower branch of the quantum capacitance signal is significantly enhanced, as indicated by the blue arrow.
The schematic energy diagram in Fig.~4d illustrates the mechanism: when the bias is below the Fermi level and lower than the ABS energy, holes preferentially tunnel into the system, increasing the occupation of the lower ABS branch and thereby showing a stronger signal in the CSD.
Similarly, when a +\SI{200}{\micro\volt} bias is applied to N2, the upper branch of the signal becomes more pronounced, as shown in Fig.~4c and highlighted by the orange arrow.
The corresponding schematic in Fig.~4e shows that, in this case, electrons more readily tunnel into the state above the Fermi level, enhancing the population of the upper ABS branch and its gate reflectometry signal.

A finer bias dependence measurement with only QD2 on resonance is shown in Fig.~4f. 
Here we measure the gate sensor response as a function of the plunger gate voltage of QD2 \var{V}{PG2} and bias voltage $V_2$. 
The two lines indicate when the QD level aligns with the ABS. 
Along the left resonance, the signal strength rises significantly below \SI{-125}{\micro V}, which can be interpreted as the point where Fermi level in N2 aligns with the QD2 and ABS energy levels. Along the right resonance, the signal is stronger above +\SI{125}{\micro V}. 
The signal strengths are plotted in the right panel in Fig.~4f.
Combining the two linecuts together, allows us to extract the ABS energy gap 2$\Delta_{ABS} \approx$ \SI{250}{\micro V}
The ABS energy can also be independently extracted from the CSDs by converting \var{V}{PG2} to an energy, through the PG2 lever arm. 
Given the lever arm of PG2 is 0.31 (see SI), and the voltage difference between the double lines is \SI{0.82}{mV}, we estimate 2$\Delta_{ABS}=\SI{254.2}{\micro V}$, which agrees well with the gap size extracted from Fig.~4f, thus supporting our interpretation for the origin of the gate-sensing signal. .

\section{Conclusions}
In this work, we have shown that gate sensing can be used to probe a Kitaev chain device where two QDs are coupled via tunable ABSs. In an open system, coupled to the normal leads, we demonstrated that gate reflectometry allows us to distinguish between ECT and CAR mediated coupling between the QDs.   
In a closed system the gate sensing signal reflects the quantum capacitance of the system, and provides information about the ground state parity. In particular we find that quasiparticle poisoning events, lead to fast switching of the ground state parity. Furthermore, we identified the normal reservoirs as a potential source of poisoning, confirmed by artificially injecting quasiparticles into the QD. In order to use gate reflectometry to study Kitaev chain based parity qubits in the future, it is important to understand in more detail the mechanisms that limit the parity lifetime in our devices and to develop protocols to use the gate sensing signal to tune up Kitaev devices in the absence of normal leads. 

\section{Acknowledgements}
We thank Di Xiao, Candice Thomas and Michael J. Manfra for providing the semiconductor heterostructures used in this work.
We thank O.W.B.~Benningshof and J.D.~Mensingh for technical assistance with the cryogenic electronics.
The research at Delft was supported by the Dutch National Science Foundation (NWO),  Microsoft Corporation Station Q and a grant from Top consortium for Knowledge and Innovation program (TKI). 
S.G. acknowledges financial support from the Horizon Europe Framework Program of the European Commission through the European Innovation Council Pathfinder grant no. 101115315 (QuKiT). 

\section{Author contributions}
I.K. fabricated the device, with input from C.G.P.. 
Y.Z. and I.K. fabricated the resonator circuits.
C.G.P. and I.K. designed the measurement set-up.
Measurements were performed by Y.Z., I.K., S.L.D.tH, and T.D..
N.v.L. and F.Z. contributed to understanding and interpretation of the data through regular discussions. 
The manuscript was written by Y.Z. and S.G., with input from all coauthors. 
S.G. supervised the experimental work in Delft.

\section{Data availability}
All raw data obtained in relation to this manuscript and the scripts to produce the figures are available on Zenodo~\cite{DataRepo}.

\bibliography{bibliography.bib} 


\end{document}


\title{Supplementary Material: Gate reflectometry in a minimal Kitaev chain device }

\maketitle

\tableofcontents
\clearpage
\section{Methods}
\subsection{Device fabrication}
The detailed device fabrication recipe is described in \cite{moehle2021insbas}. The aluminum loop is defined by lithography and wet etching, followed by the deposition of two Ti/Pd normal metallic contacts N1 and N2. A \SI{20}{nm} of AlOx dielectric layer is placed on top of the chip via atomic layer deposition (ALD) to separate electrostatic gates and the 2DEG. A Ti/Pd electrostatic depletion gate is evaporated for forming a one-dimensional channel. A Cr meander resistor (about \SI{1}{MOhm}) is attached to the depletion gate further away from the active region of the device. This resistor serves as a low-pass filter to prevent RF signal leaking to the ground. On top of the depletion gate, two more Ti/Pd gate layers are evaporated to form finger gates. There are \SI{20}{nm} of AlOx layer between each gate layer. These finger gates are used to form tunnel barriers and tune chemical potential on each section. Measurements are performed in a dilution refrigerator with a base temperature of \SI{35}{mK}.

\subsection{Device characterization}
The basic characterization is performed using DC and RF transport measurement via normal leads. 
Fig.~S1 \textbf{(d)} shows an example of the ABSs when sweeping the \var{V}{ABS} gate voltage. To test the flux dependence, an out-of-plane magnetic field \var{B}{y} is applied. 
The tunneling current shows a periodicity of \SI{27}{\micro T} and agrees well with the loop area (\SI{58}{\micro m^2}).

Coulomb diamonds are measured from both normal leads for the two QDs shown in Fig.~S1 \textbf{(f)} and \textbf{(g)}.
Superimposed dashed lines are used to estimate charging energies and the lever arm of the plunger gates.
Both dots showed similiar charging energies of \SI{1.05}{mV} and lever arm 0.3.

To perform fast RF measurements, we use a rastering method similar to the method in \cite{kulesh2025flux}. 
An arbitrary waveform generator (AWG) is used to generate sawteeth voltage and apply on the two dot plunger gates.
When measure a CSD with the size of \var{N}{1} $\cdot$ \var{N}{2} data points, one channel on the AWG sends a single sawtooth wave to one plunger gate while the other channel sends \var{N}{2} periods of waves. 
Sawteeth signals from the AWG are connected to the optical isolator, breaking any potential ground loops and are added to the constant DC offsets via a voltage divider.
The voltages are fed to the device through the filtered low-frequency lines in the fridge, as shown in Fig.~S2.

The AWG is used to trigger the sawteeth voltage output and data acquisition on the vector network analyzer (VNA). 
On the trigger event, the VNA starts to acquire \var{N}{1} $\cdot$ \var{N}{2} points in the total measurement time of \var{T}{} and stores the data in its memory.
After the full acquisition is done, the one-dimensional data array is mapped into a \var{N}{1} $\cdot$ \var{N}{2} matrix which corresponds to the CSD.

\subsection{Resonator Characterization and data processing}
We use an off-chip resonator to perform RF measurements. Each resonator contains an inductor and a parasitic capacitance (0.4 to \SI{0.6}{pF}) to ground via the bond wire. The inductors have designed values of 1500, 1200, 500 and \SI{350}{nH}, resulting in the resonance frequencies of 265, 363, 512, and \SI{889}{MHz}. All resonator signals can be measured simultaneously by multiplexing. 

The scattering parameters $S_{21}$ measured in the reflectometry circuit are complex. To better illustrate the complex signal, we plot the measured data (for example, a CSD) on the complex plane in a histogram. Coulomb blockade data contribute to the most concentrated blobs on the histogram, indicated by the black arrows. An elongated distribution of data points corresponds to the on-resonance signal in the charge stability diagram. To include the majority of information of the measurement, we can project the data to the vector that connects the Coulomb blockade and the elongated blob. Then we normalize this this new real quality and plot them as \var{\Tilde{S}}{21}~\cite{prosko2024flux}.

The reflected signal of two gate resonators can be fitted using a model for the transmission \var{S}{21} through a directional coupler reflectometry circuit~\cite{probst2015efficient,khalil2012analysis,malinowski2022radio}.
\begin{equation}
    S_{21}(f)=ae^{i\alpha}e^{-2\pi if\tau}\Big[1-\frac{(Q_i/|Q_e|)e^{i\phi}}{1+2iQ_i(f/f_r-1)}\Big]
\end{equation}
The first few factors describe the environmental effect on the reflected signal, where $a$ is the scaling factor of the amplitude, $\alpha$ is the phase shift and the elctronic delay caused by photons propagate in the cable. The second part describes a perfect resonator where $f$ denotes the probe frequency, $f_r$ the resonance frequency, $Q_i$ is the internal quality factor and $Q_e$ the external quality factor, $\phi$ quantifies the impedance mismatch. 

\subsection{QD-ABS interactions with gate sensing}
Gate-based reflectometry provides an effective approach for probing quantum dot–Andreev bound state (QD–ABS) interactions when conventional transport measurements are not available.
These interactions are essential for tuning up  Kitaev chains.
In this device, we demonstrate that gate sensing can be used to probe the ABS spectrum when the tunneling spectroscopy is not available.
The ABS energy can be tuned in two ways: by sweeping the gate voltage or by applying an out-of-plane magnetic field to introduce flux through the junction. 
A schematic of the measurement configuration, with the field direction indicated, is provided in Fig.~S5a.

When the QD level aligns with the ABS energy, particles can tunnel between the QD and the ABS, as shown in Fig.~S5b. 
The RF signal applied to the dot plunger gate slightly modulates the dot level, allowing tunneling events between the dot level and the ABS level.
As sweeping the chemical potential of the QD, the dot energy level aligns with the hole-like part and electron-like part of the ABS, illustrated in Fig.~S5b.
Quantum capacitance arises when such tunneling happens.

We observe ABS energy modulation in QD-ABS CSD in Fig.~S5c.
The spacing between each pair of double lines corresponds to the ABS energy, scaled by the lever arm factor $\alpha$. 
When the ABS merges into the bulk SC gap, tunneling between QD and the gap become much stronger and thus gives weaker gate sensing signals ($V_{ABS}$ around \SI{305}{mV}).
In addition to gate voltage tuning, we observe periodic oscillations of the double-line feature as a function of out-of-plane magnetic field, shown in Fig.~S5d.
This field response provides clear evidence that the ABS is localized within a superconducting junction, with the observed periodicity consistent with the geometric size of the loop.

\subsection{Numerical simulation}
The system in this experiment can be described as a three-site system with two QDs and a hybrid section with the following Hamiltonian~\cite{zatelli2024robust}: 

\begin{equation}
\begin{split}
    H & = \sum_{i=L,M,R} H_i + H_T\\
    H_i & = (\mu_i+E_{Zi})n_{i\uparrow} + (\mu_i-E_{Zi})n_{i\downarrow} + U_{i}n_{i\uparrow}n_{i\downarrow}+\Gamma_i(c^\dagger_{i\uparrow}c^\dagger_{i\downarrow}+c_{i\uparrow}c_{i\downarrow}),\\
    H_T & = t_L(c^\dagger_{H\uparrow}c_{L\uparrow}+c^\dagger_{H\downarrow}c_{H\downarrow})+t_{SOL}(c^\dagger_{H\downarrow}c_{L\uparrow}-c^\dagger_{H\uparrow}c_{H\downarrow})\\
    & + t_R(c^\dagger_{R\uparrow}c_{H\uparrow}+c^\dagger_{R\downarrow}c_{H\downarrow})+t_{SOR}(c^\dagger_{R\downarrow}c_{H\uparrow}-c^\dagger_{R\uparrow}c_{H\downarrow})+h.c..
\end{split}
\end{equation}
Here  $H_i$ is the Hamiltonian for each site ($i=L,R$ for the two QDs and $i=M$ for the hybrid). $n_{i\sigma}=c^{\dagger}_{i\sigma}c_{i\sigma}$ is the charge occupancy number of the state with spin $\sigma$ on site $i$. $\mu_i$ is the chemical potential energy, $E_{zi}$ is the Zeeman energy, $U_i$ is the charging energy, and $\Gamma_i$ is the induced superconducting gap. The tunneling Hamiltonian, $H_T$, describes the coupling between QDs and the hybrid region. $t_{L/R}$ and $t_{SOL/SOR}$ describe the coupling strength of the spin-conserving and spin-flipping tunneling process. The parameters used for the numerical simulations are: \var{E}{ZL}=\var{E}{ZR} = \SI{1.2}{\Gamma}, \var{E}{ZM} = 0.3 $\Gamma$ for Zeeman energy, \var{U}{L}=\var{U}{R} = \SI{7}{\Gamma}, \var{U}{M}=0 for charging energy, \var{\Gamma}{L} = \var{\Gamma}{R} = \SI{0.5}{\Gamma} for the induced gaps on the QDs, \var{t}{L,R} = \SI{0.8}{\Gamma}, \var{t}{SO,L,R} = \SI{0.4}{\Gamma} for the tunneling strength, where \var{\Gamma}{M} = $\Gamma$ is the induced gap on the ABS.

To estimate quantum capacitance signal from the model, we first calculate the charge occupation number $\langle n_i \rangle$ on each QD at odd and even ground state.
\begin{equation}
\begin{split}
     \langle n_{i,odd} \rangle =\sum_{\sigma=\uparrow,\downarrow} \bra{o}c^{\dagger}_{i\sigma}c_{i\sigma}\ket{o},\\
     \langle n_{i,even} \rangle =\sum_{\sigma=\uparrow,\downarrow} \bra{e}c^{\dagger}_{i\sigma}c_{i\sigma}\ket{e},
\end{split}
\end{equation}
Where $\ket{o}$ and $\ket{e}$ is the odd and even ground state, respectively.

Quantum capacitance $C_q$ can be qualitatively estimated by taking the numerical derivative of $\langle n_i \rangle$ with respect of the chemical potential $\mu_i$ for each site:
\begin{equation}
    C_q = \alpha e^2 \frac{d\langle n_i \rangle}{d\mu_i}
\end{equation}
We use $t_L=0.8$, $t_{SOL} = 0.32$, $t_R=0.9~t_L$, $t_{SOR}=0.9~t_{SOL}$, $E_{ZL} = E_{ZR}=1.2$, $E_{ZM}=0.3$, $U_L = U_R = 7$, $U_2=0$, $\Gamma_L = \Gamma_R =0.5$, $\Gamma_M=1$, all parameters are scaled with the induced superconducting gap $\Gamma_M$.
These parameter are used for the simulation in Fig.~1c and Fig.~3 d-f.

\newpage
\section{Supplementary Figures S1 to S5}
\begin{figure}[h]
\centering
    \includegraphics{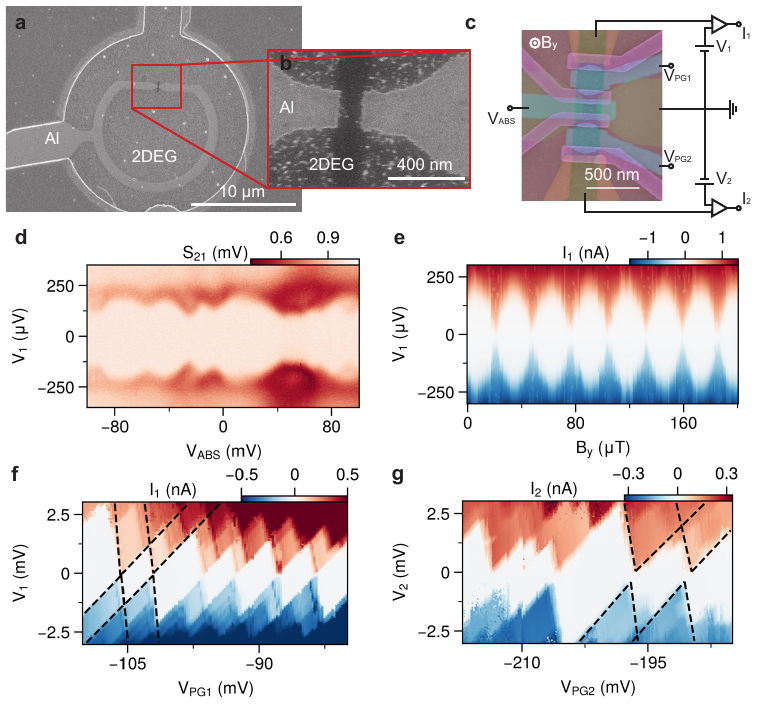}
    \caption{\textbf{Device description and basic characterizations.} 
    \fl{a}  A zoomed out SEM image of aluminum loop of a copied device that identical to the measured device. 
    \fl{b}  Zoomed in of \textbf{a} to the junction part.
    \fl{c}  A SEM image of a copied device that identical to the measured device after depositing all gate layers, together with DC transport measurement circuit. 
    \fl{d}  RF-spectroscopy of the Josephson junction as a function of the gate voltage measured from the top lead in \textbf{c}. 
    \fl{e}  Current measured from the top lead as a function of applied bias and perpendicular field $B_y$. 
    \fl{f}  Coulomb diamonds of QD1 (the top QD in \textbf{c}). The lever arm is estimated to be 0.32 by the black dashed line.
    \fl{h} Coulomb diamonds of QD2 (the bottom QD in \textbf{c}). The lever arm is estimated to be 0.31 by the black dashed line.
     }
    \label{fig:S1}
\end{figure}

\begin{figure}
\centering
    \includegraphics
    {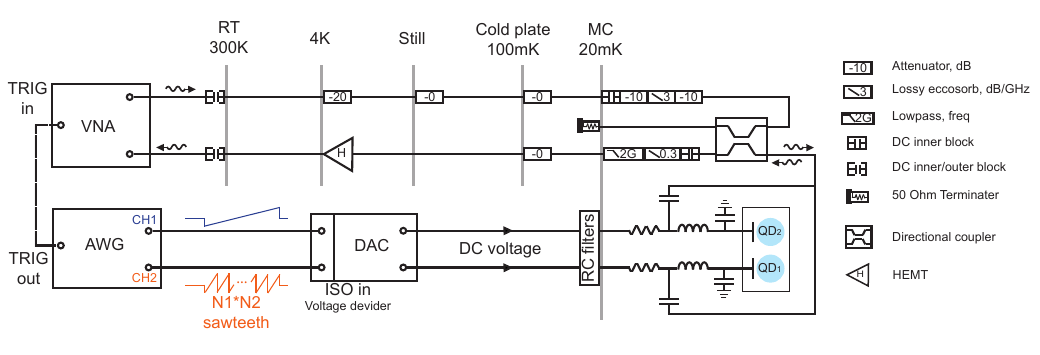}
    \caption{\textbf{Measurement setup.} A vector network analyzer (VNA) is used to perform RF measurements. RF signals pass through the microwave circuit, illustrated on the top part of the figure. An Arbitrary waveform generator (AWG) is used to apply two sawtooth voltages and add on top of the gate voltages of the two plunger gates through DC lines to the device.}
    \label{fig:S2}
\end{figure}

\begin{figure}
\centering
    \includegraphics{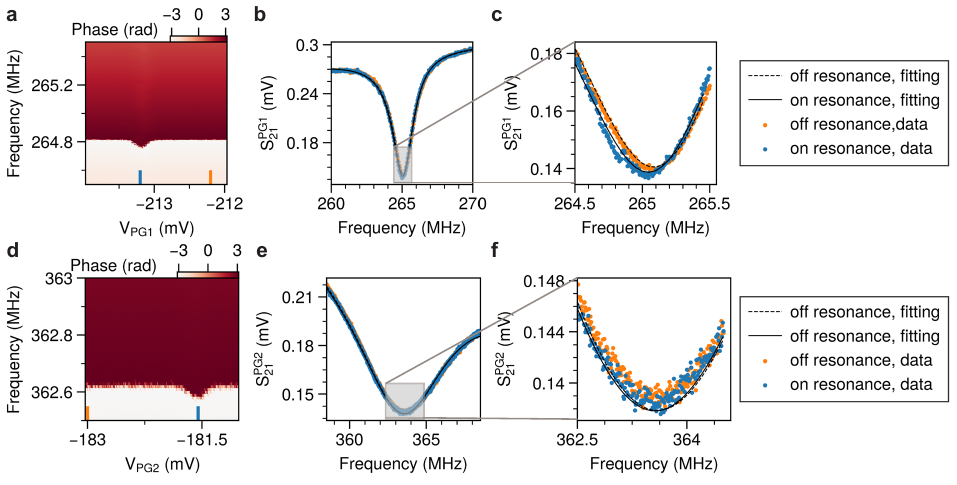}
    \caption{\textbf{Resonator fitting.} 
    \fl{a} The phase response of resonator PG1 when sweeping \var{V}{PG1}. As sweep across the coulomb resonance (blue marker), the resonance frequency slightly shifts.
    \fl{b} A zoomed out frequency scan of resonance PG1, together with the fitting results.
    \fl{c} A zoomed in frequency sweep of \textbf{b}. When dot 1 is on and off resonance, the resonance frequency slightly shifts.
    \fl{d} A similar frequency vs \var{V}{PG2} for resonator PG2. The frequency shift appears when QD2 is on resonance. 
    \fl{e} A zoomed out frequency sweep of resonator PG2.
    \fl{f} Zoomed in resonance frequency sweeps when QD2 is on and off coulomb resonance, with fitting.
   }
    \label{fig:S3}
\end{figure}

\begin{figure}[h]
\centering
    \includegraphics
    {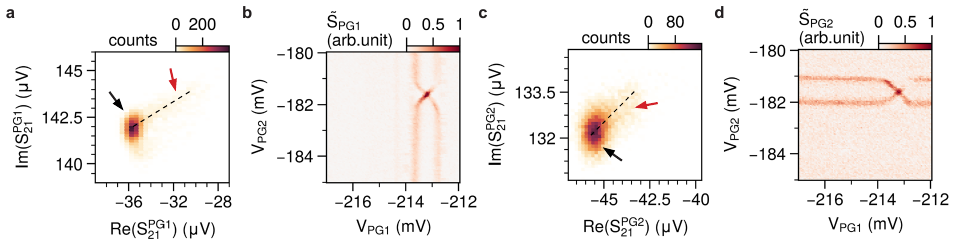}
    \caption{\textbf{Resonator signal processing.} 
    \fl{a,c} Two-dimensional histogram of complex resonator response of the measurements shown in \textbf{(b,d)}. The higher count (indicate by the black arrow) represent the signal at coulomb blockade. a faint tail (red arrow) represent the signal when the dot is on coulomb resonance.  
     }
    \label{fig:S4}
\end{figure}

\begin{figure}[h]
\centering
    \includegraphics[width = 0.48\textwidth]{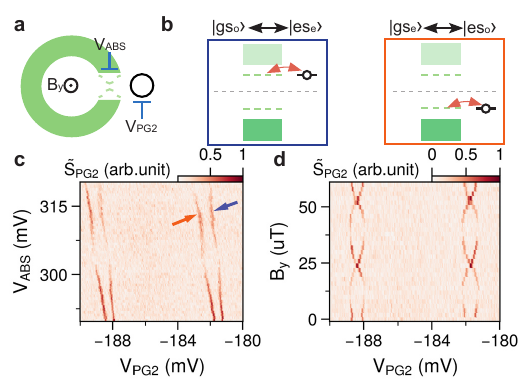}
    \caption{\textbf{Dot-ABS interactions.} 
    \fl{a} Schematic of measurement configuration. Here we only couple dot 2 to the ABS and dot 1 is off resonance. Out-of-plane magnetic field is applied to tune the flux through the loop. 
    \fl{b} Schematic of possible particle hopping in a QD-ABS coupled system. Different ground states involve in each case.
    \fl{c} QD-ABS interaction when the normal leads are decoupled. Each pair of double lines correspond of a QD resonance. As we change the ABS gate voltage, the QD resonances reflect the changes of ABS energy.
    \fl{d} Flux dependence of the dot-ABS interactions.
   }
    \label{fig:S5}
\end{figure}

\clearpage
\bibliography{bibliography}